\documentclass[
    ,final            % use final for the camera ready runs
  ]
  {aipproc}

\layoutstyle{6x9}

\newcommand{\GeV}{\ensuremath{\mathrm{Ge\kern -0.12em V}}}
\newcommand{\TeV}{\ensuremath{\mathrm{Te\kern -0.12em V}}}

\def\qq{\mathrm{q\bar{q}}}

\begin{document}

\title{Gauge Coupling Unification, SUSY Scale and Strong Coupling Running}


\author{Dimitri Bourilkov}{
  address={Physics Department, University of Florida, P.O. Box 118440,
Gainesville, FL 32611, USA}}

\begin{abstract}
The apparent unification of gauge couplings around 10$^{16}$~$\GeV$ is one
of the strong arguments in favor of Supersymmetric extensions of the
Standard Model (SM). In this contribution two new analyses of
the gauge coupling running, the latter
using in contrast to previous studies not data at the Z peak but at
LEP2 energies, are presented.
The generic SUSY scale in the more precise novel approach
is $93\ <\ M_{SUSY}\ <\ 183\ \GeV$,
easily within LHC, and possibly even within Tevatron reach.
\end{abstract}

\maketitle

%%%%%%%%%%%%%%%%%%%%%%%%%%%%%%%%%%%%%%%%%%%%
%% MAINMATTER
%%%%%%%%%%%%%%%%%%%%%%%%%%%%%%%%%%%%%%%%%%%%

\section{Introduction}

In this paper we address the hot topic of the possible scale at which
Supersymmetry could manifest itself by performing indirect analyses based
on the running of gauge couplings in two ways:
\begin{itemize}
 \item first a - by now ``traditional''~\footnote{See e.g.~\cite{Amaldi}.}
  - analysis of gauge coupling unification
  in a grand unification theory, using the latest experimental inputs, mainly at
  the Z peak scale, combined with a detailed statistical approach
 \item a novel analysis, including for the first time high precision data
  collected at LEP2, which can shed light on the actual running of couplings
  from the Z peak to $\approx$~207 $\GeV$.
\end{itemize}

In the first approach in 
contrast to the SM the Minimal Supersymmetric Standard Model (MSSM)
leads to a single unification scale of a Grand Unified Theory (GUT),
if we let the couplings run accordingly.
The relevant parameters are: $M_{SUSY}$ - a single generic SUSY scale
where the spectrum of supersymmetric particles starts to play a role, and
$M_{GUT}$ - the scale of grand unification where the electromagnetic,
weak and strong coupling come together as an unified coupling $\alpha_{GUT}$.

\vspace{-0.11cm}
\section{``Traditional'' Running Couplings Analysis}

We update the analysis~\cite{Bourilkov:2004zu}, using the same fitting
procedure and experimental inputs except for the stong coupling where we
incorporate new data.
Different measurements and world averages typically fall in two groups which
we will call ``high'' and ``low'' values:\\
$ \alpha_{s}(M_Z) = 0.1223 \pm 0.0038\ \ from\ \Gamma_h / \Gamma_{\mu}\ at\ Z\ peak\ $~\cite{LEPEWWG}\\
$ \alpha_{s}(M_Z) = 0.1182 \pm 0.0027\ \ S.Bethke\ World\ aver.\ LEP/SLC,\ Deep\ Inel.\ Scatt.$~\cite{Bethke}\\
$ \alpha_{s}(M_Z) = 0.1179 \pm 0.0030\ \ \sigma^0_l = \sigma^0_h\Gamma_l/\Gamma_h$~\cite{LEPEWWG}\\
$ \alpha_{s}(M_Z) = 0.1170 \pm 0.0026\ \ Aleph\ 4-jet\ rate$~\cite{ALEPH}.

The results of the fits are summarized in Table 1.
\begin{table}[htb]
\renewcommand{\arraystretch}{1.20}
\caption{Fit results - all for 2-loop-Renormalization-Group running.}
{\begin{tabular}{@{}lclll@{}} \hline
Inputs            & Threshold correction & $M_{SUSY}$        & $M_{GUT}$           & $1/\alpha_{GUT}$ \\
                  &    [\%]              &[$\GeV$]           &[$\GeV$]             & \\ \hline
\multicolumn{5}{c}{Low $\alpha_{s}(M_Z) = 0.117-0.118$} \\ \hline
S. Bethke 2004    &     - 4              & $10^{1.6 \pm 0.5}$& $10^{16.6 \pm 0.15}$& $22.4\pm 0.9$ \\
$\sigma^0_l = \sigma^0_h\Gamma_l/\Gamma_h$
                  &     - 4              & $10^{1.6\pm 0.55}$& $10^{16.6 \pm 0.17}$& $22.6\pm 1.0$ \\
Aleph 4-jet rate  &     - 4              & $10^{1.8 \pm 0.5}$& $10^{16.6 \pm 0.15}$& $22.8\pm 0.9$ \\ \hline
\multicolumn{5}{c}{High $\alpha_{s}(M_Z) = 0.1223$} \\ \hline
$\Gamma_h / \Gamma_{\mu}\ at\ Z\ peak$
                  &     - 4              & $10^{1.0 \pm 0.6}$& $10^{16.8 \pm 0.2}$ & $21.3\pm 1.1$ \\ \hline
\end{tabular}}
\end{table}

The strong coupling is a key for interpreting the results:
the ``high'' $\alpha_{s}(M_Z)$ values require an uncomfortably low SUSY
scale $\sim$~10 $\GeV$ -  way below the experimental lower limit
$\sim$~100 $\GeV$.
The ``low'' $\alpha_{s}(M_Z)$ values favor
values for the SUSY scale just above the present limits, e.g. for
$\alpha_{s}(M_Z)$ = 0.1182 we get $M_{SUSY} < 265\ \GeV$ at 
one-sided 95 \% CL - well in the LHC or even Tevatron direct discovery range.

\vspace{-0.11cm}
\section{Analysis of LEP2 data}

\begin{figure}[htb]
\centerline{\resizebox{0.89\textwidth}{6.2cm}{\includegraphics{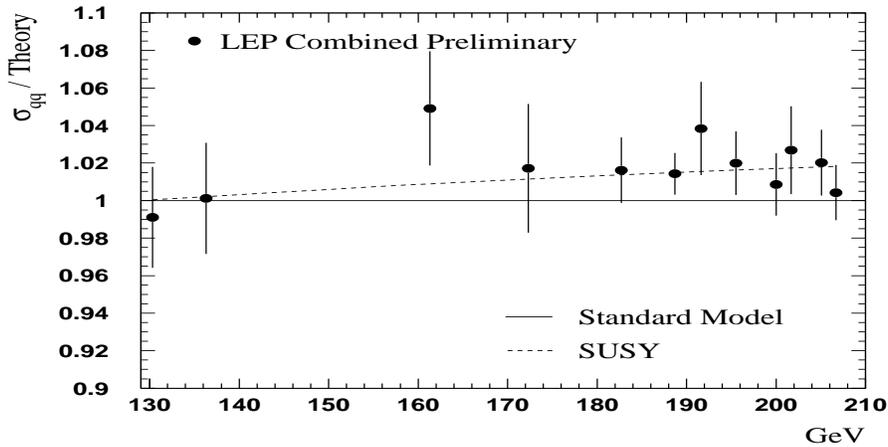}}}
%\vspace*{8pt}
\vspace*{-3pt}
\caption{LEP2 combined preliminary $\sigma_{\qq}$ measurements and the
results from SM or SUSY fits.}
\label{fig:qqfit}
\end{figure}

\begin{figure}[!Hhtb]
  \centering
  \begin{tabular}{cc}
    \resizebox{0.48\textwidth}{6.2cm}{\includegraphics{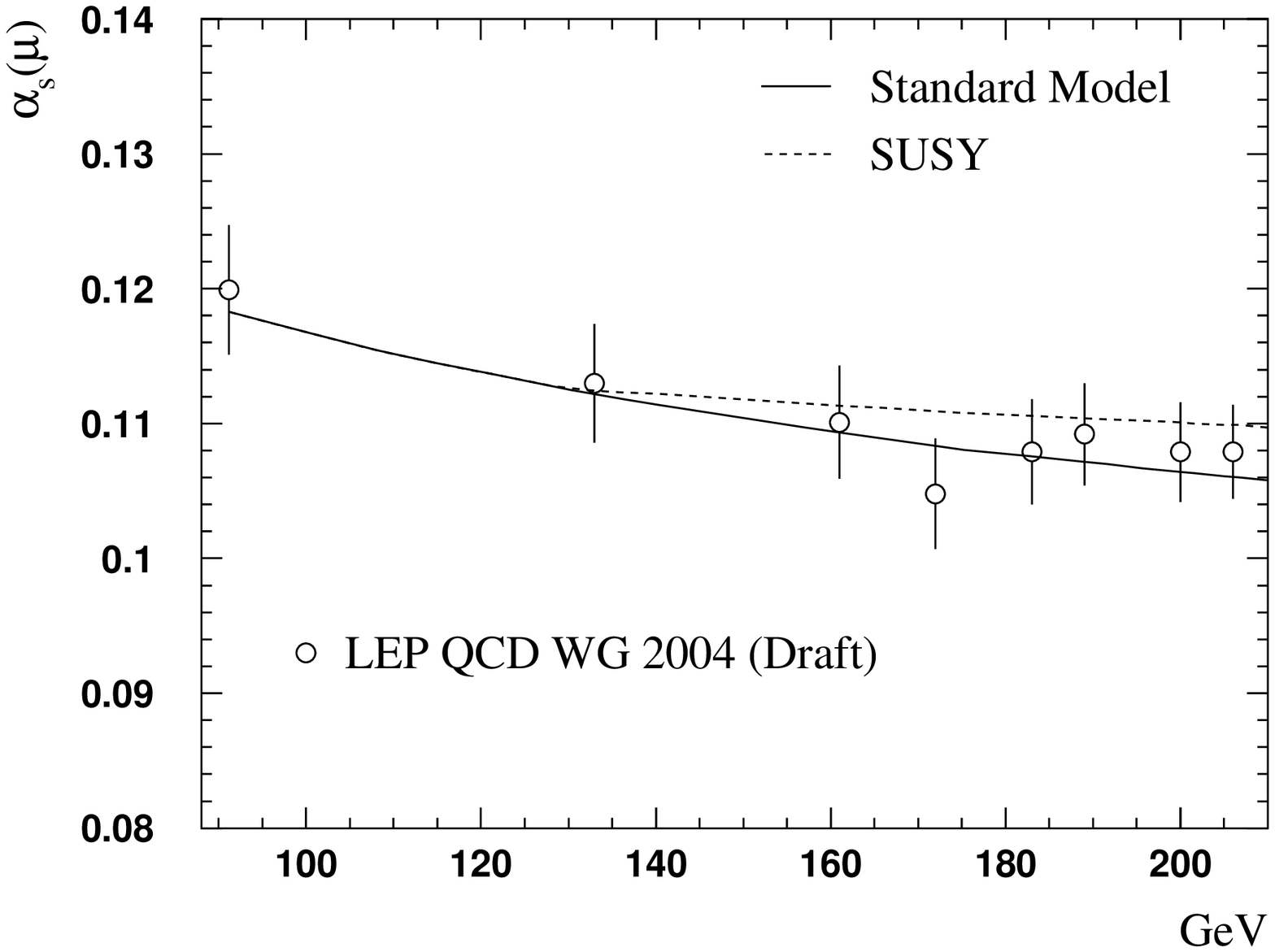}}
    \resizebox{0.48\textwidth}{6.2cm}{\includegraphics{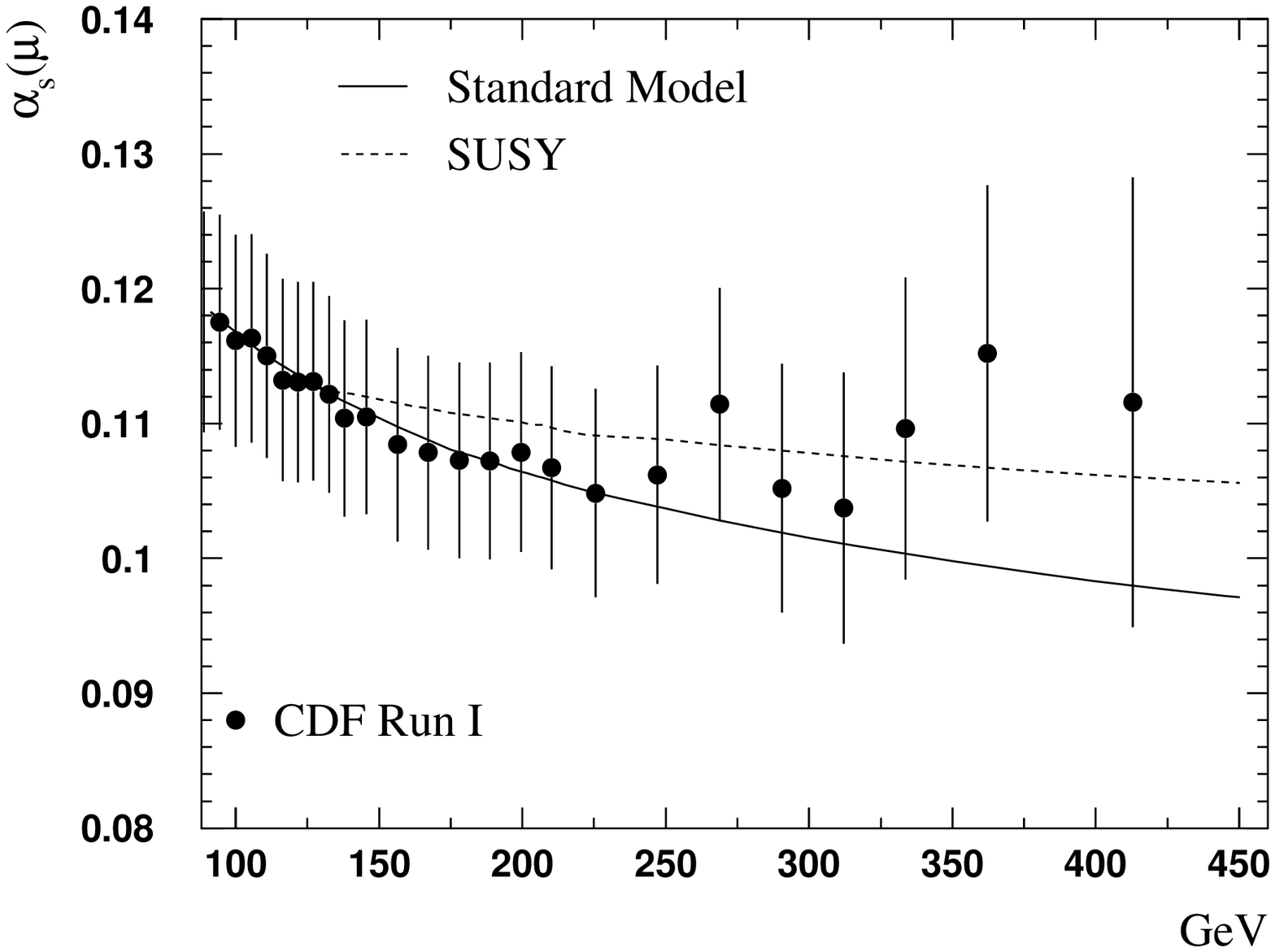}}
  \end{tabular}
\vspace*{-3pt}
\caption{Measurements of $\alpha_s$ above the Z pole together with the SM
or SUSY predictions, using our best fit for the latter.}
\label{fig:alfas}
\end{figure}

We perform a new more ``direct'' analysis by adding more data. LEP2 has measured the
$\qq$ cross section above the Z pole~\cite{LEPEWWG} with combined
precision from the four experiments $\sim$~1~\% and theoretical
uncertainty $<$~0.3~\%. We fit the data computing the running of the
couplings from the Z peak to the highest LEP2 energies in the SM or in
MSSM with the scale as free parameter. All three couplings change in
a different way in the two cases if the SUSY scale is in this mass range.
As all couplings affect the theory prediction for $\sigma_{\qq}$, they
could give measurable effects at this level of precision.
The slight ``excess'' $\sim$~2 standard deviations is ``filled'' nicely
if the couplings run as predicted in MSSM. Practically all high precision
points are slightly above the SM prediction as could be the case if SUSY
starts to take over - see Figure~\ref{fig:qqfit}. Of course a more mundane
explanation, which can not be ruled out at the current level of precision
and understanding, is that we have a normalization problem either on the
theoretical or experimental side.
The scale is fitted to be
$$93\ <\ M_{SUSY}\ <\ 183\ \GeV\ \ \ \ at\ 95\ \% \ confidence\ level.$$
Precision measurements of the strong coupling $\alpha_s$ above the Z pole
can provide new information. This coupling changes the fastest in
this energy range, but is also the hardest to measure. Systematic uncertainties
are dominant both at LEP2 (theory and hadronization) and 
CDF (energy scale and resolution; interplay with the gluon parton density
function uncertainty at high X values). As shown
in Figure~\ref{fig:alfas} the experimental precision starts to approach the 
effects predicted by using our best fit value for the SUSY scale from
$\sigma_{\qq}$. If the Tevatron experiments benefit fully from the increased
statistics in Run 2 to control the systematics better and probe higher energy
scales, they could possibly provide new independent evidence for
the scale of supersymmetry.

\begin{thebibliography}{0}

\bibitem{Amaldi}
U.~Amaldi et al., Phys. Lett. {\bf B260}, 447 (1991).

%\cite{Bourilkov:2004zu}
\bibitem{Bourilkov:2004zu}
  D.~Bourilkov,
  %``A fresh look at gauge coupling unification,''
  Int.\ J.\ Mod.\ Phys.\ A {\bf 20} (2005) 3328
  [arXiv:hep-ph/0410350].
  %%CITATION = HEP-PH 0410350;%%

\bibitem{LEPEWWG}
LEP Electroweak Working Group, Summer 2005 update, hep-ex/0511027.

\bibitem{Bethke}
S.~Bethke, hep-ex/0407021.

\bibitem{ALEPH}
ALEPH Collaboration, Eur.Phys.J. {\bf C27}(2003),1. The total error used
follows the discussion in hep-ex/0211012.

\end{thebibliography}
\end{document}